# No stable gravitationally or electrostatically bound atoms in n-space for n > 3

Mario Rabinowitz

Armor Research, 715 Lakemead Way, Redwood City, CA 94062-3922

E-mail: Mario715@earthlink.net

**Abstract**

It is demonstrated in general that stable gravitational or electrostatic orbits are not possible for spatial dimensions n ≥ 4. It is thus shown that atoms cannot exist and that planetary motion is not possible in higher dimensional space. Furthermore, angular momentum cannot be quantized in the usual manner in 4-space, leading to interesting constraints on mass. Thus Kaluza Klein and string theory may be impacted since it appears that the unfurled higher dimensions of string theory will not permit the existence of energetically stable atoms. This also has bearing on the search for deviations from $1/r^2$ of the gravitational force at sub-millimeter distances. The results here imply that such a deviation must occur at less than ~ $10^{-8}$ cm, since atoms would be unstable if the curled up dimensions were larger than this.

## 1. Introduction

A framework combining hierarchy theory (Dirac 1937, 1938) and string theory was proposed by postulating the existence of 2 or more compact dimensions in addition to the standard 3 spatial dimensions that we commonly experience (Argyres, Dimopoulos, and March-Russell, 1998). In this view, gravity is strong on a scale with the higher-dimensional compacted space, and only manifests itself as being weak on a macroscopic 3-dimensional scale. One prediction (Arkani-Hamed et al, 1998) is that if there are only 2 compacted dimensions of radius $r_c \sim 10^{-2}$ cm, it should be possible to detect a deviation of the Newtonian $1/r^2$ force law at this scale. It will be shown in this paper that for $r_c \sim 10^{-8}$ cm, common electrostatically bound atoms will not be stable. For convenience some previously derived results for gravitationally bound atoms (Rabinowitz, 1990, 2001) will be used.

## 2. No Energetically Bound Circular Orbits for n > 3 in n-space

Gravitational and electrostatic long-range attractive forces can be expressed in n-space $n = 3, 4, 5, ...$ , as
$$F_n = \frac{-K_n}{r_n^{n-1}}. \tag{2.1}$$

For the gravitational force (Rabinowitz, 2001)
$$K_{Gn} = \frac{2\pi G_n Mm\Gamma(n/2)}{\pi^{n/2}}, \tag{2.2}$$

where we will consider the orbiting mass $m \ll M$. For the electrostatic force
$$K_{En} = \frac{2\pi R_{En} Qq\Gamma(n/2)}{4\pi\varepsilon\pi^{n/2}}, \tag{2.3}$$

where a body of mass m with negative charge q orbits around a positive charge Q. $R_{En}$ is a model dependent factor that relates the electrical force in n-space to the electrical force in 3-space, and $\varepsilon$ is the permittivity of free space.

Equating $F_n$ to the centripetal force, yields the kinetic energy:
$$\frac{-K_n}{r_n^{n-1}} = \frac{-mv_n^2}{r_n} \Rightarrow \tfrac{1}{2}mv_n^2 = \frac{K_n}{2r_n^{n-2}}. \tag{2.4}$$

The potential energy is
$$V_n = -\int \vec{F}_n \bullet d\vec{r} = \frac{-K_n}{(n-2)r_n^{n-2}}. \tag{2.5}$$

Adding eqs. (2.4) and (2.5) gives the total energy
$$\begin{aligned}E_n &= \tfrac{1}{2}mv_n^2 + V_n = \frac{K_n}{2r_n^{n-2}} + \frac{-K_n}{(n-2)r_n^{n-2}} \\ &= \left[\tfrac{1}{2} - \frac{1}{(n-2)}\right]\frac{K_n}{r_n^{n-2}} = \left[\frac{n-4}{n-2}\right]\frac{K_n}{2r_n^{n-2}}\end{aligned} \tag{2.6}$$

The total energy $E_n \leq 0$ for $n \geq 4$. This result applies both classically and quantum mechanically since quantization has not yet been invoked, and quantization will not change the sign of the co-factor $K_n / r_n^{n-2}$. Therefore there are no energetically bound circular orbits for n > 3 in n-space. We will next consider non-circular quantized orbits.

## 3. Non-Circular Orbits in Higher Dimensions

In higher dimensional space central force trajectories are generally neither circular, nor elliptical, as the orbits become non-closed curves. In fact elliptical orbits occur only for potentials $\propto 1/r$ and $\propto r$. Although only circular orbits have been considered so far, the more complicated central force problem where there is also a radial velocity, yields the same conclusion regarding the instability of atoms for $n \geq 4$. For non-circular orbits, we shall take into consideration the effective potential energy as illustrated in Fig. 1. The general case can be put in the form of a one-dimensional radial problem in terms of the effective potential energy of the system,

$$V_n' = V_n + L^2 / 2mr_n^2. \tag{3.1}$$

where $V_n(r)$ is the potential energy of the system, and L is the angular momentum which remains constant because there are no torques in central force motion.

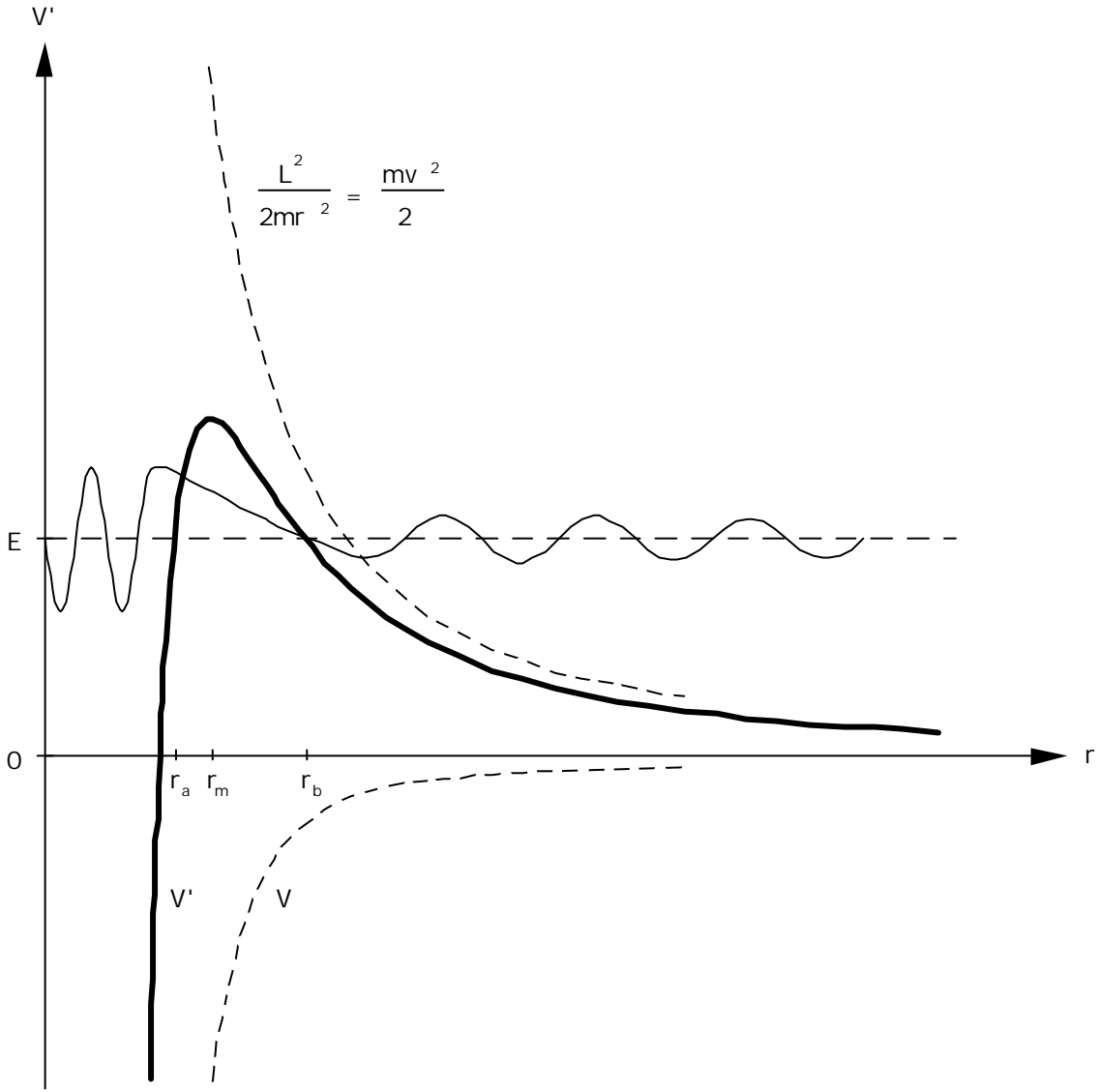

Figure 1. Effective potential energy in n-space with maxium value at $r_m$, showing tunneling through the finite barrier of width $r_b - r_a$ at total energy E.

The orbits are not energetically bound if $E_n - V'_n(r_m) \geq 0$, where $r_m$ is the radius of the circular orbit at the maximum of $V'_n$ (cf. Fig. 1). Those orbits for which $0 < E_n < V'_n(r_m)$ are classically, but not quantum mechanically bound. If atoms could be formed in this region, they would be only metastable since the finite width of the potential energy barrier presented by $V_n'$ permits the orbiting body to tunnel out. Let us see if energetically bound atoms can even be formed.

The general equation of motion that includes radial motion is

$$F_n = \frac{-K_n}{r_n^{n-1}} = m\frac{d^2 r_n}{dt^2} - \frac{L^2}{mr_n^3}. \tag{3.2}$$

Let us substitute eq. (2.5) for the potential energy into eq. (3.1) for the effective potential energy to determine if there is an n that satisfies:

$$E_n(r_m) - V_n'(r_m) = E_n + \frac{K_n}{(n-2)r_m^{n-2}} - \frac{L^2}{2mr_m^2} \geq 0. \tag{3.3}$$

The maximum value of $V_n'$ occurs at $r_m$, and is obtained by setting $dV_n'/dr = 0$. This is the same as dropping the radial force term $m\ddot{r}_n$ in the force eq. (3.2):

$$\frac{-K_n}{r_m^{n-1}} = \frac{-L^2}{mr_m^3} \Rightarrow r_m = \left[\frac{mK_n}{L^2}\right]^{1/(n-4)}. \tag{3.4}$$

This is the radius $r_m$ for a circular orbit at the maximum value of $V_n'$. Trajectories with $r > r_m$ are unbound both classically and quantum mechanically as can be seen clearly from Fig. 1. Substituting for $E_n$ from eq. (2.6) into eq. (3.3),

$$\left[\frac{n-4}{n-2}\right]\frac{K_n}{2r_m^{n-2}} + \frac{K_n}{(n-2)r_m^{n-2}} - \frac{L^2}{2mr_m^2} \quad 0. \tag{3.5}$$

Combining the first two terms, and substituting eq.(3.4) into eq. (3.5):

$$\frac{K_n}{r_m^{n-2}} \geq \frac{L^2}{2m}r_m^{-2} \Rightarrow 1 \geq \frac{L^2}{2mK_n}\left(r_m^{n-4}\right) = \frac{L^2}{2mK_n}\left(\left[\frac{mK_n}{L^2}\right]^{1/(n-4)}\right)^{n-4} = 1. \tag{3.6}$$

Eq. (3.6) implies that the circular orbit at $r = r_m$ is at the highest energy state, and thus

$$E_n(r_m) = V_n'(r_m) > E_n(r_n) \tag{3.7}$$

Let us first look at $E_n$ non-relativistically by means of the uncertainty principle with $p \sim \Delta p \sim \hbar/2\Delta x$, and $r \sim \Delta x$:

$$E_n \sim \frac{(\Delta p)^2}{2m} - \frac{K_n}{(n-2)(\Delta x)^{n-2}} = \frac{\hbar^2}{8mr^2} - \frac{K_n}{(n-2)r^{n-2}}. \tag{3.8}$$

Classical orbits can exist in the region $0 < E_n < V_n'(r_m)$. However, since they would be subject to quantum tunneling, classical orbits would only be metastable. For n ≥ 4 and r small enough to make $E_n < 0$, the orbiting body would spiral in to r = 0 both quantum mechanically and classically since then the negative potential energy term dominates in eq. (3.8). For large kinetic energies, this needs to be checked relativistically.

Let us look at $E_n$ by means of the uncertainty principle with $r \sim \Delta x$, and the relativistic energy equation :

$$E_n = \left[(pc)^2 + m_o^2 c^4\right]^{1/2} + \frac{-K_n}{(n-2)r_n^{n-2}}$$

$$\sim \left[\left(\frac{\hbar}{2r_n}\right)^2 c^2 + m_o^2 c^4\right]^{1/2} + \frac{-K_n}{(n-2)r_n^{n-2}} \tag{3.9}$$

Eq. (3.9) indicates that for n ≥ 4 and r small enough to make $E_n < 0$, the orbiting body would spiral in to r = 0 both quantum mechanically and classically since then the negative potential energy term dominates in eq. (3.9).

Therefore for n ≥ 4, quantum orbits of any configuration are not energetically bound. Classical and quantum orbits can exist in the region $0 < E_n < V_n'(r_m)$. However, since they would be subject to quantum tunneling, these orbits would only be metastable. For n ≥ 4 and r small enough to make $E_n < 0$, the orbiting body would spiral in to r = 0 both quantum mechanically and classically since then the negative potential energy term dominates in eqs. (3.8) and (3.9).

## 4. Quantization of Angular Momentum in 4-Space

In all dimensions except in 4-space, the dependence of angular momentum, L, on $r_n$ allows the orbital radius to adjust in the quantization of L. This and no binding energy for atoms for ≥ 4-space has ramifications for the 4-space Kaluza-Klein unification of general relativity and electromagnetism, as well as for string theory. Let us briefly examine the ramifications of the quantization of L, without quantization of r, in 4-space for gravitational and electrostatic atoms

Equating the gravitational force (Rabinowitz, 2001) to the centripetal force in 4-space for circular orbits of a two-body gravitationally bound atom of reduced mass $\mu = mM/(m+M)$:

$$F_{Gn} = \frac{-2\pi G_n Mm \Gamma(n/2)}{\pi^{n/2} r_n^{n-1}} \xrightarrow{n=4} \frac{-2\pi G_4 Mm}{\pi^2 r_4^3} = \frac{-2 R_G G_3 Mm}{\pi r_4^3} = -\mu \frac{v_4^2}{r_4}, \quad (4.1)$$

where $G \equiv G_3 = G_4/R_G$. $R_G$ is a model dependent factor that relates the gravitational force in n-space to the gravitational force in 3-space. Similarly for $R_E$ and the electrical force. Solving eq. (4.1) for the angular momentum, $L_G$, of the two-body gravitational atom, and quantizing $L_G$:

$$L_G = \mu v_4 r_4 = [2\mu R_G G_3 Mm/\pi]^{1/2} = j\hbar, \quad (4.2)$$

Equating the electrostatic force (mks units) to the centripetal force in 4-space for a two-body electrostatically bound atom:

$$F_{En} = \frac{2\pi R_{En} Q(-q) \Gamma(n/2)}{4\pi\varepsilon \pi^{n/2} r_n^{n-1}} \xrightarrow{n=4} -\left(\frac{e^2}{4\pi\varepsilon}\right) \frac{2R_E}{\pi r_4^3} = -(\alpha\hbar c)\frac{2R_E}{\pi r_4^3} = -\mu \frac{v_4^2}{r_4}, \quad (4.3)$$

where $\varepsilon$ is the permittivity of free space, $\alpha \approx 1/137$ is the fine structure constant, and the electronic charge $e = Q = q$. Solving eq. (4.3) for the angular momentum, $L_E$, of the two-body electrostatic atom, and quantizing $L_E$:

$$L_E = \mu v_4 r_4 = [2\mu\alpha\hbar c R_E/\pi]^{1/2} = j\hbar. \quad (4.4)$$

Quantization lets us set $L_E = L_G$, since they are both $= j\hbar$. Assuming $R_G = R_E$, this yields a condition on the product of the two masses in terms of the Planck mass $M_P$,

$$Mm = \alpha\left(\frac{\hbar c}{G}\right) = \alpha M_P^2. \quad (4.5)$$

This says that the gravitational angular momentum in 4-space can only be quantized if the product of the two masses $Mm = \alpha M_P^2 \approx M_P^2/137$. Empirically, the electron mass can be related to $\alpha$ and the proton mass,

$$m_e = 10.22\alpha^2 M_p \Rightarrow M_p m_e = 10.22\alpha^2 M_p^2. \quad (4.6)$$

It is an interesting coincidence that not only does $\alpha$ enter into eqs. (4.5) and (4.6), but that they can be put into a somewhat similar form, where an extra factor of $10\alpha$ takes us from the macroscopic to the subatomic domain.

Eqs. (4.2) and (4.4) imply quantization of products and sums of the masses if $R_G$ and $R_E$ are not quantized. Eq. (4.2) implies

$$\frac{(Mm)^2}{M+m} \propto (j\hbar)^2. \qquad (4.7)$$

Eq. (4.4) implies

$$\frac{Mm}{M+m} \propto (j\hbar)^2. \qquad (4.8)$$

## 5. Discussion

Except for the $s = 0$ state, identically the same results in 3-space are obtained for the Bohr-Sommerfeld semi-classical approach as from the Schroedinger equation. Though the latter is done by the more difficult route of solving this second order differential equation with associated Laguerre polynomials. Therefore it is reasonable to expect the same results in higher dimensions. Even if they were to differ, there is no question that the orbiting mass could tunnel out of the finite width effective potential energy barrier. So in general, higher dimensional atoms are not stable.

A framework has been proposed for unifying the weak gravitational force with the strong force by postulating the existence of 2 or more compact dimensions in addition to the standard 3 spatial dimensions that we commonly experience. In this view, gravity is strong on a scale with higher-dimensional compacted space, and only manifests itself as being weak on a larger 3-dimensional scale.

Although modern hierarchy theory is independent of string theory, it borrows from and has much in common with string theory. It does not require the (9 spatial + 1 time) dimensions of string theory. It utilizes the same concepts of restricting other forces that reside inside the compacted dimensions to remain therein, while allowing the gravitational force to manifest itself from the compressed space into 3-space. A testable prediction of one version of this theory is that if there are two and only two additional dimensions there should be a deviation from the $1/r^2$ Newtonian force at sub-millimeter dimensions (Arkani-Hamed et al, 1998). As shown by eq. (2.1), in a 5-dimensional space, one may expect a $1/r^4$ dependence of the gravitational force.

The degree of arbitrariness in this hierarchy theory can be illustrated by its prediction of the size of the extra compacted dimensions

$$r_c \sim 10^{\frac{30}{d}-17} \text{ cm,} \qquad (5.1)$$

where d = n - 3. For d = 1 (4-space), eq. (4.1) predicts $r_c \sim 10^{13}$ cm $\sim 10^8$ miles. The distance of the earth to the sun is $9.3 \times 10^7$ miles. So there cannot be only one extra dimension, since the Newtonian gravitational force is well established at this scale. For d = 2 (5-space), both extra dimensions would have $r_c \sim 10^{-2}$ cm. For d = 3 (6-space), the three extra dimensions would all be at the atomic dimension $r_c \sim 10^{-7}$ cm. The 6 extra dimensions of string theory would all have $r_c \sim 10^{-12}$ cm, so the impact on gravity would be at the nuclear scale.

The conclusions of hierarchy/string theory of a sub-millimeter compaction size do not appear to be compelling. The predictions regarding the size of the compacted dimensions can be modified down to the Planck length of $10^{-35}$ m, if experiment shows no deviation from standard Newtonian gravity at larger sizes. So far no deviation has been found down to 0.15 mm -- almost ruling out the 5-dimensional space predictions given by eq. (5.1). The extra dimensions are confined by branes. Until now the size of branes seemed to be so small that they would not contradict experimental findings since other forces have been probed to sub-nuclear sizes. The investigation for a deviation from Newtonian gravity is spurred by allowing branes to be $\sim 10^{-1}$ mm in radius.

## 6. Conclusion

It has been shown that neither gravitational nor electrostatic quantized orbits are stable for spatial dimensions n ≥ 4. Even though classical orbits can exist in the region $0 < E_n < V_n^{'}(r_m)$, they would only be metastable as they would be subject to tunneling through the effective potential energy barrier which has a finite width. Thus the findings here indicate that it is highly unlikely that a deviation of the $1/r^2$ gravitational force law will be found at the sub-millimeter scale, or atoms would not be stable.

## Acknowledgment

I wish to thank Felipe G. Garcia for helpful discussions, and for help with the figure.

## References

Argyres, P.C., Dimopoulos, S., and March-Russell, J. (1998). *Phys.Lett.* **B441**, 96.
Arkani-Hamed, N., Dimopoulos, S., and Dvali, G. (1998) *Phys.Lett.* **B429**, 263.
Dirac, P.A.M. (1937). *Nature* **139**, 323.
Dirac, P.A.M. (1938). *Proc. Royal Soc. London* **A165**, 199.
Rabinowitz, M. (1990) *IEEE Power Engineering Review* **10**, No. 4, 27.
Rabinowitz, M. (2001) *Int'l Journal of Theoretical Physics*, **40**, 875.